\documentclass[twocolumn, twocolappendix]{aastex631}
\usepackage{amsmath}

\begin{document}

\title{Two 100 TeV Neutrinos Coincident with the Seyfert Galaxy NGC~7469}

\correspondingauthor{Giacomo Sommani}
\email{sommani@astro.ruhr-uni-bochum.de}

\author[0000-0002-0094-826X]{Giacomo Sommani}
\affiliation{Ruhr-Universität Bochum, Fakultät für Physik und Astronomie, \\
Astronomisches Institut (AIRUB), D-44780 Bochum, Germany}

\author[0000-0002-5605-2219]{Anna Franckowiak}
\affiliation{Ruhr-Universität Bochum, Fakultät für Physik und Astronomie, \\
Astronomisches Institut (AIRUB), D-44780 Bochum, Germany}


\author[0000-0002-1460-3369]{Massimiliano Lincetto}
\affiliation{Ruhr-Universität Bochum, Fakultät für Physik und Astronomie, \\
Astronomisches Institut (AIRUB), D-44780 Bochum, Germany}

\author[0000-0001-8206-5956]{Ralf-Jürgen Dettmar}
\affiliation{Ruhr-Universität Bochum, Fakultät für Physik und Astronomie, \\
Astronomisches Institut (AIRUB), D-44780 Bochum, Germany}




\begin{abstract}

In 2013, the IceCube collaboration announced the detection of a diffuse high-energy astrophysical neutrino flux. The origin of this flux is still largely unknown. The most significant individual source is the close-by Seyfert galaxy NGC\,1068 at 4.2-sigma level with a soft spectral index. To identify sources based on their counterpart, IceCube releases realtime alerts corresponding to neutrinos with a high probability of astrophysical origin. We report here the spatial coincidence of two neutrino alerts, IC220424A and IC230416A, with the Seyfert galaxy NGC\,7469 at a distance of 70\,Mpc.
We evaluate, a-posteriori, the chance probability of such a coincidence and discuss this source as a potential neutrino emitter based on its multi-wavelength properties and in comparison to NGC 1068 by performing a Goodness-of-Fit test. The test statistic is derived from a likelihood ratio that includes the neutrino angular uncertainty and the source distance. We apply this test first to a catalog of AGN sources and second to a catalog of Seyfert galaxies only.
Our a-posteriori evaluation excludes the possibility of an accidental spatial coincidence of both neutrinos with the Seyfert galaxy NGC\,7469 at 3.2-sigma level, leaving open the possibility that either one or both neutrinos originated from the source.
To be compatible with non-detections of TeV neutrinos, the source would need to have a hard spectral index.

\end{abstract}



\section{Introduction} \label{sec:intro}

The IceCube Neutrino Observatory detected a diffuse astrophysical neutrino flux above 30\,TeV in 2013 \citep{neutrinoflux}. However, the origin for most of this flux is still largely unknown. In general, neutrino production requires acceleration of protons or heavier nuclei to high energies. In the presence of a proton or photon target, those high-energy particles can interact to produce pions. The charged pions produce neutrinos in their decay chain. Active galaxies belong to the most promising candidates for high-energy neutrino production~\citep[see][for a recent review]{Kurahashi:2022utm}.

Several individual neutrino source candidates were identified, including the flaring $\gamma$-ray blazar TXS 0506+056 \citep{txs2018,txs2018excess} and the tidal disruption events AT2019dsg~\citep{Stein:2020xhk} and AT2019fdr~\citep{Reusch:2021ztx}. The most significant source candidate is the nearby Seyfert galaxy NGC 1068 with a significance of $4.2\sigma$~\citep{ngc1068neutrinos}. The neutrino emission from NGC 1068 is best described by a power-law with a spectral index of $3.2\pm 0.2$. The excess consists of $79^{+22}_{-20}$ neutrinos mostly at TeV energies~\citep{ngc1068neutrinos}. Neutrino production in NGC 1068 can be explained by several models involving coronae, shocks, winds, and their interaction with dense material~\citep[see e.g.,][]{2022arXiv220702097I,Murase:2022dog,2022ApJ...939...43E,Inoue:2019yfs}. Further evidence for neutrino production by Seyfert galaxies was presented by \citet{IceCube:2023jds} and \citet{2023arXiv230609018N}, who identified NGC 4151 and NGC 3079 as additional potential neutrino emitters. 

Here, we present the detection of two 100\,TeV neutrinos, IC220424A and IC230416A, selected by the IceCube realtime system, spatially coincident with the Seyfert galaxy NGC\,7469 located at a redshift of $0.016$.
The distance of NGC\,7469 is not well constrained, and thus we assume a value of 70\,Mpc in a standard $\Lambda$CDM cosmology in agreement with most of the recent literature, in particular with the GOALS project~\citep{2009PASP..121..559A} which has studied NGC\,7469 extensively as part of the JWST Early Release Science program~\citep{jwst_early}.
This work presents a statistical test to evaluate the chance coincidence of the two neutrinos and NGC 7469. We test two source catalogs, the Million Quasars~\citep{milliquas} and the Turin-SyCAT catalog~\citep{turincatalog}.

The work is structured as follows: We first summarize the multi-wavelength properties of NGC\,7469 and compare it to the neutrino source candidate NGC\,1068 in Sec.~\ref{sec:ngc7469}. Section~\ref{sec:catalog} and~\ref{sec:neutrinodata} present the source catalogs and neutrino dataset, respectively. Our method to calculate the chance coincidence is outlined in Sec.~\ref{sec:test} and the results are outlined in Sec.~\ref{sec:results}.
In Sec.~\ref{sec:discussion} we report a discussion of our findings.

\section{NGC~7469}
\label{sec:ngc7469}

NGC\,7469 is an active galaxy of the Seyfert 1 class and hosts also starburst activity in a circum-nuclear ring~\citep{2023ApJ...953L...9Z}. 
This prominent Seyfert galaxy is well studied in all wavelengths regimes and considered a prime candidate for high energy particle emission. Together with NGC\,1068 it belongs to the original sample of six galaxies studied by \citet{1943ApJ....97...28S}. With an infra-red luminosity of $10^{11.65} L_\odot$ \citep{2009PASP..121..559A}, it is a luminous (but not ultraluminous) infrared galaxy.
The mass of its supermassive black hole was estimated to 1-2\,$\times10^7 M_\odot$ through reverberation mapping~\citep{2014ApJ...795..149P} 
and gas kinematics from a high angular resolution ALMA study~\citep{alma_ngc7469}. The latter also constrains the inclination of the central gas disk to $\sim11\degr$.
The source has shown X-ray variability on time scales $<$ 1 day~\citep{Nandra_1998}, suggesting an origin of the emission in the inner regions of the AGN.
From high angular resolution studies in the radio-continuum with the Multi-Element Radio Linked Interferometer Network (MERLIN) \citep{alberdi_paper, corejet_7469} and with Very Long Baseline Interferomtry (VLBI) observations taken with the Very Long Baseline Array (VLBA) \citep{lonsdale_paper}, a core-jet-like structure on a 100\,pc scale is inferred.

\subsection{Comparison to NGC~1068}
NGC\,7469 is five-times further away than NGC\,1068, which is located at 14.4\,Mpc~\citep{Koss:2022qwh}. Both sources host an active nucleus and starburst activity. Different from NGC\,7469, NGC\,1068 is a Seyfert 1.9 galaxy~\citep{Koss:2022qwh}, i.e. our view onto the nucleus is partly obscured by the dust torus. 
The inclination of the inner disk of NGC\,1068 is measured to be 40\degr - 41\degr~\citep{inclination_1068}.
This explains that the observed X-ray fluxes of both sources are comparable despite the difference in distance. 
After correcting for the absorption, the intrinsic X-ray flux in the energy range of $14-195$ keV of NGC\,1068 is $\sim3$ times larger than that of NGC\,7469~\citep{BASSxray}.

A comparison of the spectral energy distribution of NGC\,7469 and NGC\,1068 is shown in Fig.~\ref{fig:sed}.

\section{Source Catalogs}
\label{sec:catalog}

Our chance coincidence calculation (see Sec.~\ref{sec:test}) depends critically on the number of sources we consider as neutrino emitters and how we assume that the neutrino flux scales with their properties, which enters as a weight in our analysis. Since our calculation is done a-posteriori, we use several catalogs to avoid a fine-tuning of the significance calculation. We consider the following two source catalogs:
\begin{itemize}
    \item All the 351 sources in the Turin-SyCAT catalog~\citep{turincatalog}, a multifrequency catalog of Seyfert galaxies. 
    This choice is motivated by the recent evidence of neutrino emission from the Seyfert galaxy NGC~1068~\citep{ngc1068neutrinos}. The same catalog was used by~\citet{2023arXiv230609018N}.
    \item All the 71345 Active Galactic Nuclei (AGNs) in the Milliquas catalog detected in X-ray (including quasars)~\citep{milliquas}.
    In general, the processes suggested for neutrino production in NGC~1068, could also take place in other types of AGN.
    Therefore, to be more agnostic, we test a catalog of all AGN.
    As several models suggest the X-ray flux as good tracer for neutrino emission~\citep{Inouexray, kotaxray, alixray}, we limit our search to the X-ray detected AGNs.
\end{itemize}
In both cases, we use the inverse of the source distance squared as a weight~(Sec.~\ref{subsubsec:signal}). 
This overly simplified standard candle assumption does most properly not describe the reality, but is a conservative assumption given the lack of knowledge on the neutrino production and its multiwavelength tracers.
A weighting scheme based on a more precise model, which takes into account the individual source properties, should result in a better description of the relative weights, and thus lead to a rejection of the null hypothesis at higher significance.
The redshift provided by the corresponding catalog was used as a distance estimator.
We chose our test to be agnostic, and therefore did not assume any correlation of the neutrino flux with the flux in a given band.
We do note that several models suggest the intrinsic X-ray flux as a good tracer for neutrino emission \citep{Inouexray, kotaxray, alixray}. However, estimating the intrinsic X-ray is model dependent \citep{BASSxray}. 
As we desire to design a statistical test whose results are robust with a minimal set of assumptions, we chose to use only the distance as a tracer of the expected neutrino flux (see Sec.~\ref{sec:test}).
However, for completeness reasons, we also performed a test using the intrinsic fluxes from \citet{BASSxray} as tracer in Appendix~\ref{sec:xraytest} and, as alternative agnostic method, another test setting equal weights to each source in~Appendix~\ref{sec:noweight}.

\section{Neutrino Data}
\label{sec:neutrinodata}

This work is based on neutrino detections by the IceCube neutrino observatory, which consists of 5160 optical sensors embedded in 1\,km$^3$ of the Antarctic ice sheet close to the Amundsen-Scott South Pole Station \citep{Aartsen_2017}. 
IceCube detects neutrino interactions with the surrounding ice or nearby bedrock through detection of Cherenkov radiation from charged secondary particles.
The light-emission signatures can be classified in two types of event morphologies: tracks and cascades.
Track-like events are produced by muons, which originate from charged-current (CC) interactions of muon neutrinos or from cosmic-ray showers.
Cascade-like events can result from neutral-current (NC) interactions of all-flavor neutrinos, or from CC interactions of electron and tau neutrinos.
While the angular reconstruction of cascade-like events has large uncertainties of $\sim10^\circ$, the direction of track-like events can be reconstructed with an uncertainty of less than 1$^\circ$ \citep{Aartsen_2014}. Therefore, most searches for neutrino sources rely on track-like events. Track-like events are further classified into starting and through-going track events, based on the location of the neutrino interaction being inside or outside the instrumented detector volume.

\subsection{IceCube Realtime Program}\label{subsec:realtime}

To find neutrino source candidates, the IceCube Realtime System was established in 2016 \citep{realtime_established} and updated in 2019 \citep{Blaufuss:20199c}. It selects high-energy neutrino events based on their signalness, i.e. their probability to be of astrophysical origin~\citep{realtime_established}. As of 2019, the through-going track events, are divided into two streams, called ``Gold" and ``Bronze" \citep{Blaufuss:20199c}, with an average signalness of 30$\%$ and 50$\%$ respectively. Each realtime alert is initially released as a General Coordinates Network (GCN) Notice\footnote{\url{https://gcn.nasa.gov/notices}}, and updated after a few hours as a GCN Circular\footnote{\url{https://gcn.nasa.gov/circulars}} \citep{Blaufuss:20199c}. The GCN Notice reports the direction and angular uncertainty derived with the reconstruction algorithm SplineMPE \citep{Abbasi_2021,sommani2023}. The GCN Circular contains the results of the more sophisticated and computing-intensive reconstruction algorithm Millipede \citep{Aartsen_2014,gualda2021,sommani2023}. SplineMPE is significantly faster and therefore implemented for the first estimate of the direction. Millipede makes use of a more realistic description of the muon-track light emission that requires much more computational resources \citep{Aartsen_2014,gualda2021,sommani2023}. 

Nevertheless, recent studies based on Monte Carlo data \citep{gualda2021,sommani2023} indicate that this improved description does not necessarily result in an improved angular reconstruction, mainly because of systematic uncertainties represented by our limited knowledge of the south-pole ice. In fact, despite its simplicity, the reconstruction algorithm SplineMPE results in a more precise angular localization and is more robust against known systematic errors in Monte Carlo studies \citep{sommani2023}. 
Moreover, the initial GCN Notices (SplineMPE) have on average initial contours more than 10 times smaller in area than the GCN Circulars (Millipede). 
For the reasons outlined above, we test here the set of alert data provided by the SplineMPE algorithm (i.e. the first GCN Notice related to each event). For the set, we rely on the bronze and gold alerts released from June 2019 until October 2023.
Those realtime alerts cover a period of 4 years. 
One alert, which has a GCN Circular, is absent in this dataset.
The event is IC210503 \citep{2021GCN.29951....1I}.
According to the GCN Circular, IceCube was in a test run configuration and therefore no automated alert was circulated via GCN Notice.
Because of the absence of a GCN Notice, for the estimation of the neutrino energy of IC210503 we looked at the IceCat catalog \citep{abbasi2023icecat1}.

Our chance probability calculation (see Sec.~\ref{sec:test}) relies on the neutrino angular uncertainty, which we model as a bivariate symmetric Gaussian resembling the point-spread function for each individual event. SplineMPE provides a circularized estimate (i.e. a radius) for the 50\% and 90\% uncertainty. More precisely, we use the 50\% radius and directly translate it to the standard deviation, $\sigma$, of the bivariate Gaussian\footnote{We prefer the 50\% over the 90\% contour because the former is the one that was studied to calibrate the contours, whereas the 90\% is only scaled with a fixed factor \citep{Blaufuss:20199c}.}.


\subsection{Neutrino Doublet}
\label{sec:doublet}

\begin{figure}
\plotone{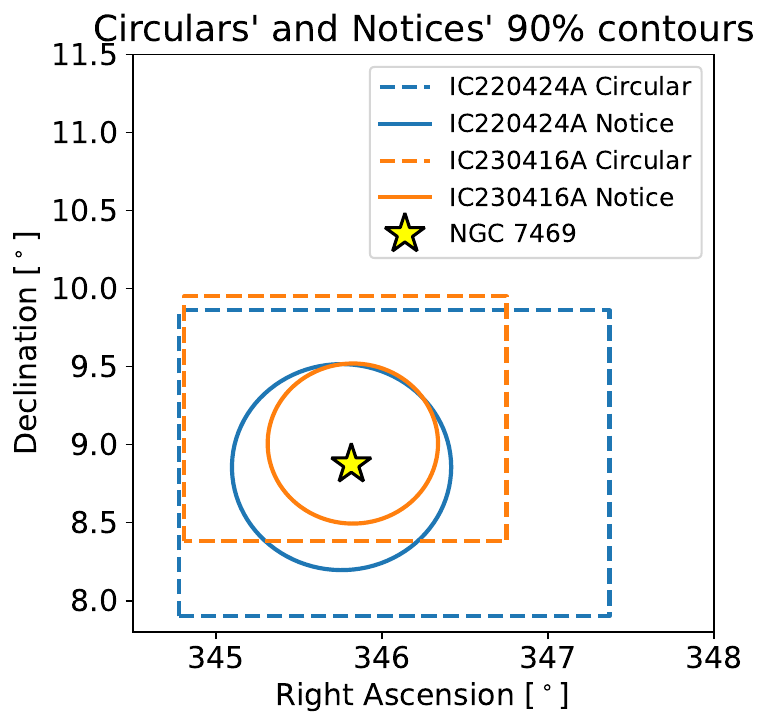}
\caption{Angular reconstructions reported in the GCN Circulars (dashed lines) and in the GCN Notices (continuous lines) for the neutrino alerts IC220424A and IC230416A \citep{2022GCN.31942....1I,2023GCN.33633....1I}. The figures show the 90$\%$ uncertainties, the best-fit directions, and the position of NGC 7469.}\label{fig:circular}
\end{figure}

On April 24, 2022, and one year later, on April 16, 2023, IceCube sent out two GCN Notices reporting the detection of the neutrino events IC220424A\footnote{GCN Notice for IC220424A: \url{https://gcn.gsfc.nasa.gov/notices_amon_g_b/136565_2186969.amon}} and IC230416A\footnote{GCN Notice for IC230416A: \url{https://gcn.gsfc.nasa.gov/notices_amon_g_b/137840_57034692.amon}}, respectively. 
The two GCN Notices were then followed by the respective GCN Circulars \citep{2022GCN.31942....1I,2023GCN.33633....1I}. Both the SplineMPE and the Millipede directions of these two alerts are compatible with the position of the Seyfert galaxy NGC~7469 at RA:~23h~03m~15.61s and DEC:~+08d~52m~26s~(J2000 Equinox)~\citep{turincatalog}, see Fig.~\ref{fig:circular}.

IC220424A was classified as a gold alert with a signalness of 50\% and an energy of 184\,TeV. IC230416A was classified as a bronze alert with a signalness of 34\% and an energy of 127\,TeV.

The neutrino directions were close to the Sun and the Moon, hampering prompt electromagnetic follow-up observations (see Sec.~\ref{subsec:followup}).
IC220424A's best-fit direction was 37.68 deg from the Sun and 22.14 deg from the Moon. 
IC230416A's best-fit direction was 44.54 deg from the Sun and 43.73 deg from the Moon.
All neutrino alert information are summarized in Table~\ref{table:nudata}.

\begin{splitdeluxetable*}{ccccccBccccBcccccc}\label{table:nudata}
\tabletypesize{\small}
\tablewidth{0pt} 
\tablecaption{Data related to the two neutrino alerts IC220424A (first line) and IC230416A (second line)}
\tablehead{
\multicolumn{6}{c}{General info} & \multicolumn{4}{c}{GCN Circular} & \multicolumn{6}{c}{GCN Notice}\\
Alert ID&Stream&Energy$^*$ [TeV]&Signalness&Moon distance [deg]&Sun distance [deg]&RA$^+_-$ [deg] (90\% error)&DEC$^+_-$ [deg] (90\% error)&2D-Gaussian's $\sigma^{**}$ [deg]&NGC~7469 distance [deg]&RA [deg]&DEC [deg]&Err 90\% [deg]&Err 50\% [deg]&2D-Gaussian's $\sigma^{**}$ [deg]&NGC~7469 distance [deg]
} 
\colnumbers
\startdata 
IC220424A&Gold&184&50\%&43.73&44.54&346.11$^{1.26}_{1.33}$&$+8.91^{0.95}_{1.01}$&0.64&0.29&345.76&+8.86&0.66&0.26&0.22&0.06\\
IC230416A&Bronze&127&34\%&22.14&37.68&345.85$^{0.90}_{1.04}$&$+9.41^{0.81}_{0.76}$&0.49&0.27&345.82&+9.01&0.51&0.20&0.17&0.13\\
\enddata
\tablecomments{$*$Most likely energy of the neutrino deduced from the parameters of the alert event under the astrophysical neutrino hypothesis. The spectrum of the diffuse astrophysical neutrino flux is assumed to be a power law with spectral index -2.19 \citep{Blaufuss:20199c}.\\
$**$Calculated as explained in Sec.~\ref{subsec:realtime} (GCN Notice) and in Appendix~\ref{sec:mill_gof} (GCN Circular).}
\end{splitdeluxetable*}

\subsection{Electromagnetic and low-energy neutrino follow-up}\label{subsec:followup}
IC220424A was followed up by \textit{Fermi}-LAT with no significant ($>$ 5 sigma) new excess emission above an energy of 100 MeV \citep{IC220424AFermiLAT}.
The alert was also followed up in the optical by MASTER-Net with no detection and an upper limit of 14.9 mag \citep{IC220424AMASTER-Net}.
In addition, a search for $\mathcal{O}$(TeV) neutrinos was performed by IceCube. No significant excess was found. The search had a sensitivity to neutrino point sources with an $E^{-2.5}$ spectrum ($E^2 \, dN/dE$ at 1 TeV) of $0.13$ GeV cm$^{-2}$ in a 1000 second time window and of $0.15$ GeV cm$^{-2}$ in a 2 days time window. 90\% of the events that IceCube would have detected in this search with the energy spectrum $E^{-2.5}$ would have energies in the approximate range between 200\,GeV and 100\,TeV \citep{IC220424AIceCubeFRA}.

IC230416A was followed up by \textit{Fermi}-LAT with no significant ($>$ 5 sigma) new excess emission above an energy of 100 MeV \citep{IC230416AFermiLAT}.
Also in this case, a search for additional $\mathcal{O}$(TeV) neutrino events was performed, and no significant excess was found. The sensitivity was the same as for the search for additional neutrinos from IC220424A \citep{IC230416AIceCubeFRA}.

\section{Estimate of Chance Coincidence}
\label{sec:test}

To evaluate if the spatial coincidence between the two neutrino alerts IC220424A and IC230416A and the Seyfert galaxy NGC\,7469 is a chance coincidence, we performed a Goodness-of-Fit test for the hypothesis that the alerts do not originate from any source in the catalog. 
This test makes use of a test statistic (Sect.~\ref{subsec:teststatistic}) calculated for neutrino doublets coincident with a source. 
Given a set of neutrino alerts, it selects the doublet with the highest test statistic and then evaluates the corresponding p-value by comparing the result with the distribution of the test statistic under the hypothesis of alerts not produced by any source (Sect.~\ref{sec:results}). 
The outcome of this test consists of a p-value expressing how likely such a coincidence is to occur from randomly-distributed neutrino alerts.

This test relies on the choice of the catalog and of a weight applied to each source in the catalog~(Sec.~\ref{sec:catalog}).
The better the weight describes the reality, the better we can reject the null hypothesis. 
We chose a simplistic model to assign a weight to all sources based only on the distance of the source.
We do not aim to test neutrino production from a population of galaxies, but rather at evaluating the chance coincidence for the neutrino doublet from one source of the catalog.

\subsection{Test statistic}\label{subsec:teststatistic}
\
The test statistic for neutrino doublets is derived from the log-likelihood ratio of:
\begin{itemize}
    \item The null hypothesis $H_0$: the doublet was not produced by any source in the catalog;
    \item The alternative hypothesis $H_1$: the doublet has been produced by the source $S$.
\end{itemize}
\
The log-likelihood ratio depends on the probability density functions (pdfs) for the doublet under the two different hypotheses.

Our proposed Goodness-of-Fit test evaluates the null hypothesis against any alternative hypothesis.
However, in constructing our test statistic, we select the alternative hypothesis that is most distinct from the scenario where neutrinos do not originate from any cataloged source.
Specifically, this alternative assumes that both neutrinos originate from a source within the catalog.
The case where one neutrino originates from a source and the other does not is more difficult to distinguish from a sample of neutrinos not produced by any cataloged source, and is discussed in further detail in Appendix~\ref{sec:unbias}.



Sect. \ref{subsubsec:null} shows the pdf under $H_0$, while Sect. \ref{subsubsec:signal} discusses the pdf under $H_1$. Finally, Sect. \ref{subsec:choices} discusses analysis choices.

\subsubsection{Probability density function under the null hypothesis}\label{subsubsec:null}
\
In the following, the i-th neutrino alert will be indicated as $A_i$. In case the neutrinos did not originate from any source in the catalog, two neutrino alerts $A_i$ and $A_j$ can be treated independently. Therefore, the probability density function for the doublet under $H_0$ is equal to the product of the pdfs for the single events:
\begin{equation}\label{eq:p0indip}
    p_0(A_i,A_j) = p_0(A_i)p_0(A_j)\,.
\end{equation}
\
In the case where the two neutrinos are not produced any source in the catalog, the pdf under the null hypothesis describes how likely it is to find an alert $A_i$ at a precise declination $\theta_i$ and energy $E_i$ by random chance.
To understand how these quantities are distributed, we consider all the alerts from IceCat-1, the IceCube Event Catalog of Alert Tracks~\citep{abbasi2023icecat1}.
IceCat-1 also reports an effective area $A_\mathrm{eff}$ binned in energy and declination.
By following the same binning of the effective area, we count how many neutrinos from the catalog fall into each bin. From this, we derive a ratio $r_k(\theta_i, E_i)$ that indicates the fraction of events in the bin k where $\theta_i$ and $E_i$ are found\footnote{Some bins contain no events and have a ratio equal to zero. 
During the scrambling (Sec.~\ref{sec:results}), this is not a problem because it changes only the right ascension of the alerts.
It can become a problem with the injection of alerts at a source position (Appendix~\ref{sec:unbias}) or if the declinations are changed (Appendix~\ref{sec:decvar}).
In these cases, we restrict the selection to include only alerts with a declination and energy whithin a non-empty bin.}.
We assume a constant energy flux in each bin (i.e., a neutrino flux proportional to $E^{-2}$) and an isotropic distribution (i.e.,~$\propto\cos{\theta}$, accounting for the spherical geometry).
Considering normalizing factors for each bin, the null hypothesis probability for the single neutrino event is given by
\begin{equation}
    p_0(A_i) = \frac{1}{2\pi}\cos{\theta_i}E_i^{-2}\xi_k\zeta_kr_k(\theta_i,E_i)\,,
\end{equation}
where $1/(2\pi)$ accounts for the uniform distribution in right ascension, $\xi_k=E_k^+E_k^-/(E_k^+-E_k^-)$ normalizes the energy for bin k where the i-th alert is located in ($E_k^+$ and $E_k^-$ are the upper and lower energy bounds for the bin), and $\zeta_k=(\sin{\theta_k^+}-\sin{\theta_k^-})^{-1}$ normalizes the declination for the same bin ($\theta_k^+$ and $\theta_k^-$ are the upper and lower declination bounds for the bin).

The null hypothesis probability for the doublet is:
\begin{equation}\label{eq:p02nu}
     p_0(A_i,A_j)=\frac{\cos{\theta_i}\cos{\theta_j}}{\left(2\pi E_iE_j\right)^{2}}\xi_k\zeta_kr_k(\theta_i,E_i)\xi_l\zeta_lr_l(\theta_j,E_j)\, .
\end{equation}

\subsubsection{Probability density function under the alternative hypothesis}\label{subsubsec:signal}
\
Under the alternative hypothesis (i.e. the doublet with the highest test statistic has been produced by a source $S$), the pdf for a neutrino doublet produced by a specific source can be divided in two components:
\begin{itemize}
    \item A flux component $p_f$ that expresses the probability of detecting \textit{at least} two neutrinos from the source;
    \item A spatial component $p_a$ which takes into account the angular distance between the source and the neutrinos.
\end{itemize}
\ 
The flux component consists of the cumulative density function between 2 and $+\infty$ detections of a Poisson distribution:
\begin{equation}\label{eq:pf}
    p_f = \sum_{k=2}^{+\infty} \frac{\mu_S^k}{k!}e^{-\mu_S} = 1 - \sum_{k=0}^{1} \frac{\mu_S^k}{k!}e^{-\mu_S} = 1 - \left(1 + \mu_S\right)e^{-\mu_S}\,,
\end{equation}
where $\mu_S$ is the expected number of detected neutrinos from a specific source $S$.
The expected number of neutrinos depends on three components:
\begin{itemize}
    \item The neutrino flux of the source at Earth, $\mu_f(E,z_S)$, which depends on the distance of the source (hence, on the redshift $z_S$) and its neutrino energy spectrum. We treat the sources as emitting neutrinos following a power-law spectral shape with spectral index of $\gamma=2$, i.e. $\phi(E)=\phi_0 \left(\frac{E}{E_0}\right)^{-2}$, where $\phi_0 = \phi(E_0)$ is the flux normalization at source, and $E_0$ is the lower bound in the effective-area energy range from \citet{abbasi2023icecat1}.
    \item The effective area $A_\mathrm{eff}(E,\theta_S)$ that describes the properties of the detector, which depends on the neutrino energy and its declination.
    \item The duration of the experiment $T$. The longer the experiment is, the higher is the probability of detecting neutrinos from the source (we treat the sources as steady neutrino emitters).
\end{itemize}
\
The total number of expected neutrinos is given by the integral over the energy:
\begin{equation}\label{eq:muestimation}
    \mu_S = \mu(\theta_S, z_S) = T\int{\mu_f(E,z_S)\,A_\mathrm{eff}(\theta_S,E)\,dE}\,,
\end{equation}
where $\theta_S$ and $z_S$ are the declination and the redshift of the source. 
For the flux normalization it is necessary to make a choice that will be discussed in Sec.~\ref{subsec:choices}.

By integrating Eq.~\ref{eq:muestimation} assuming a linear distance-redshift relation the expected number of neutrinos from the source~$S$ is
\begin{equation}\label{eq:expectednumnu}
    \mu\left(\theta_S, z_S\right) = \frac{H_0^2 T}{4\pi z_S^2 c^2}\phi_0E_0^2\sum_kA_\mathrm{eff}\left(\theta_S,E_k\right)\frac{E_k^+ - E_k^-}{E_k^+ E_k^-}\,,
\end{equation}
where $H_0$ is the Hubble constant, $E_k$ is the energy of the $k$-th energy bin of the effective area, $E_k^-$ is its lower bound and $E_k^+$ is its upper bound.
$\mu_S$ depends on the flux normalization at source $\phi_0$, and this is an input that we need to indicate before performing the test.
If $\phi_0$ is small enough, we have $\mu_S\ll1$, i.e. on average the neutrino flux is too small to produce a detection. 
In this regime, the flux component $p_f$ (Eq.~\ref{eq:pf}) can be approximated as
\begin{equation}\label{eq:approx}
    p_f\simeq\frac{1}{2}\mu_S^2\,.
\end{equation}

For the spatial component of the pdf $p_a$ we assume a Gaussian distribution of the events\footnote{Here we use the small-angle approximation, which is particularly valid near the equator, where most of IceCube's alerts are gathered \citep{abbasi2023icecat1}.}:
\begin{equation}\label{eq:spatialcomp}
    p_a(A_i,A_j\,|\,S) = \frac{1}{4\pi^2\sigma_i^2\sigma_j^2}\exp{\left[-\left(\frac{\Omega_{Si}^2}{2\sigma_i^2}+\frac{\Omega_{Sj}^2}{2\sigma_j^2}\right)\right]}\,,
\end{equation}
where $\sigma_i$ is obtained as described in Sect. \ref{subsec:realtime} and where $\Omega_{Si}^2=\left(\varphi_S-\varphi_i\right)^2+\left(\theta_S-\theta_i\right)^2$ with $\varphi_S$ and $\theta_S$ the right ascension and the declination of the source.

From Eqs. \ref{eq:pf} and \ref{eq:spatialcomp} we have the pdf for the doublet in the signal hypothesis:
\begin{eqnarray}\label{eq:p1S}
    p_1(A_i,A_j \, | \, S) = \qquad\qquad\qquad\qquad\qquad\qquad\qquad\qquad\\\nonumber =\frac{1 - \left(1 + \mu_S\right)e^{-\mu_S}}{4\pi^2\sigma_i^2\sigma_j^2}\exp{\left[-\left(\frac{\Omega_{Si}^2}{2\sigma_i^2}+\frac{\Omega_{Sj}^2}{2\sigma_j^2}\right)\right]}\,,
\end{eqnarray}
with $\mu_S$ the expected number of neutrinos from the source $S$ from Eq. \ref{eq:expectednumnu}.
The pdf in Eq. \ref{eq:p1S} depends on the considered source $S$. 
To have a probability that depends only on the doublet, we choose the source $S$ that maximizes $p_1$.
\begin{equation}\label{eq:p1}
    p_1(A_i,A_j) = \max_S{p_1(A_i,A_j\, |\, S)}\,.
\end{equation}
\
The test statistic is determined by the likelihood ratio $\lambda(A_i,A_j)$:
\begin{equation}\label{eq:likelihoodratio}
    \lambda(A_i,A_j)=2\log{\frac{p_1(A_i,A_j)}{p_0(A_i,A_j)}}\,.
\end{equation}
\
Neglecting constant factors, using Eqs. \ref{eq:p02nu}, \ref{eq:p1S} and \ref{eq:p1}, the test statistic $TS(A_i,A_j)$ is
\begin{multline}\label{eq:ts}
    TS(A_i,A_j) = \\ = \max_S{\left\{\log{\left[1-\left(1+\mu_S\right)e^{-\mu_S}\right]}-\left(\frac{\Omega_{Si}^2}{2\sigma_i^2}+\frac{\Omega_{Sj}^2}{2\sigma_j^2}\right)\right\}} - \\
     - 2\log{\left(\sigma_i\sigma_j\right)} - \log{\left[\cos{\theta_i}\cos{\theta_j}\right]} + 2\log{\left(E_iE_j\right)} - \\ - \log{\left[\xi_k\zeta_kr_k\left(\theta_i,E_i\right)\xi_l\zeta_lr_l\left(\theta_k,E_k\right)\right]}\,.
\end{multline}

\subsection{Analysis Choices}\label{subsec:choices}
In order to apply the statistical test outlined above, we have to make a few choices. 

Our neutrino sample is described in Sec.~\ref{subsec:realtime}. We perform our tests using the alert data from the initial GCN Notices (i.e. reconstructed by the algorithm SplineMPE). This consists of 4 years of data (113 events, all Gold and Bronze alerts with the first GCN Notice realized in realtime starting from June 19, 2019, until October 4, 2023).
In addition, we have to select a source catalog. We test the two catalogs described in Sec.~\ref{sec:catalog}.

\begin{figure}
\plotone{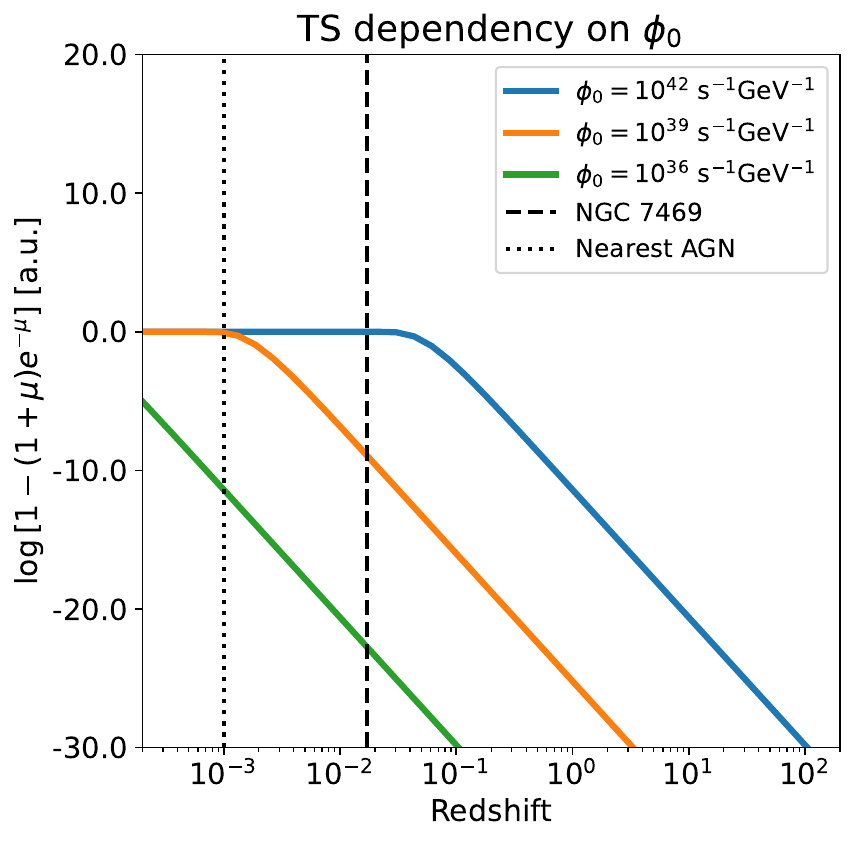}
\caption{Dependence of the test statistic on the redshift, for three different values of the neutrino flux normalization at source $\phi_0$ (at energy $E_0$).
The green contribution ($\phi_0 = 10^{36}$ s$^{-1}$GeV$^{-1}$) is in the regime in which $\mu_S\ll1$ for all sources, and makes the ordering independent of the neutrino flux normalization $\phi_0$. In this plot, the sources are taken at declination $\theta_S = +8.5$ deg.}\label{fig:flux}
\end{figure}

In Fig.~\ref{fig:flux} we show how the test statistic would be influenced by various choices of the neutrino flux normalization at source~$\phi_0$ (which is included in the estimation of $\mu_S$, see Eq.~\ref{eq:expectednumnu}).
The figure shows the component $\log{\left[1-\left(1+\mu_S\right)e^{-\mu_S}\right]}$ of the test statistic as a function of the redshift. Note that this is the only component of the test statistic which depends on redshift.
Three different choices of $\phi_0$ are investigated. 
For the case of $\phi_0 = 10^{36}$ $\mu_S\ll1$ for all sources.
In this regime, the component of the test statistic becomes $\log{\left[1-\left(1+\mu_S\right)e^{-\mu_S}\right]}\simeq2\log{\mu_S}-\log{2}$ (see~Eq.~\ref{eq:approx}) and $\mu_S\propto\phi_0$~(Eq.~\ref{eq:expectednumnu}). In this regime $\phi_0$, which, in our simplified weighting scheme, is the same for all sources, becomes just an additional constant factor in the test statistic. As a result, the ordering of the sources is independent of $\phi_0$ for all relevant redshifts and all sources are weighted with their inverse distance squared (green curve in Fig.~\ref{fig:flux}).

For larger values of $\phi_0$, $\mu_S\ll1$ only applies starting from a certain redshift $z_\mathrm{cut}$. This results in an independence of redshift at low redshift. Only distant sources will be penalized according to their inverse distance squared.
If we do not restrict ourselves to the regime in which $\mu_S\ll1$, the choice of a particular neutrino flux normalization $\phi_0$ corresponds to the choice of a specific redshift $z_\mathrm{cut}$ at which the TS begins to penalize the sources because of distance.
Different redshifts $z_\mathrm{cut}$ result in different outcomes for our statistical test.
In the regime of $\mu_S\ll1$, our test will be independent of the exact value of $\phi_0$. 
To set the test statistic in this regime, we choose $\phi_0<10^{37}$ s$^{-1}$GeV$^{-1}$.

Our weighting scheme assigns smaller weight to distant sources, but otherwise ignores the intrinsic properties of the different sources in the catalog.
Such properties could be the geometry of the source, the mass of the central supermassive black hole and the accretion rate and other (unknown) properties, which could influence the neutrino production.
Developing a more detailed description of the neutrino emission by all sources in the catalog is out of the scope of this paper. \citet{xrayseyferticecube} searched for the collective neutrino signal of Seyfert galaxies following the disk-corona model, but found no excess. However, in a simple catalog search (agnostic to any model prediction), two additional neutrino source candidates appeared, which indicates that the neutrino emission does not follow the prediction by the assumed model.
Hence, here we chose a simplified scaling, which will penalize the probability of the alternative hypothesis.
This is not a problem for our test, as long as the scenario of alerts that did not originate from any source in the catalog is even less likely than our simplified alternative scenario.
We expect both the null hypothesis and the alternative scenario to be unlikely, given our simplistic source scaling.
However, what is relevant for our test, is that the measured test statistic is more compatible with the expected test-statistic distribution of the alternative scenario compared to the scenario of alerts not produced by any source in the catalog. We verify the sensitivity of our test by studying mock neutrino alert samples with a signal doublet injected on a source of the catalog (see~Appendix~\ref{sec:unbias}).


\section{Results}\label{sec:results}
\begin{figure}
\plotone{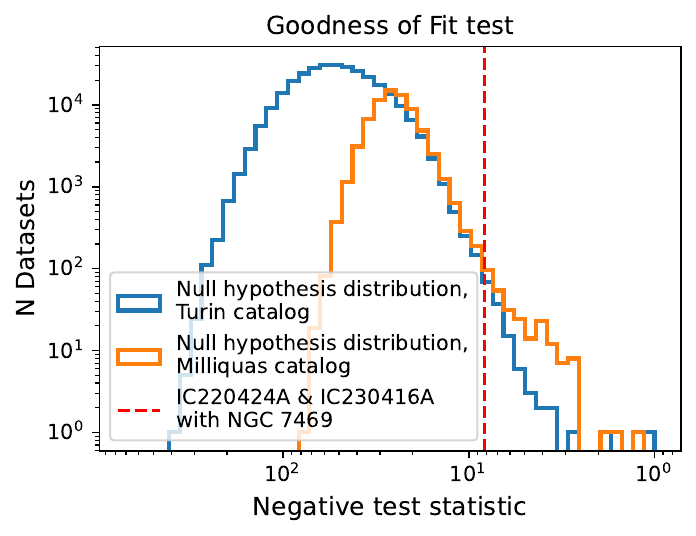}
\caption{
Test statistic distributions under the hypothesis of alerts that did not originate from any source in the Turin \citep{turincatalog} and Milliquas \citep{milliquas} catalogs.
For the Turin catalog $3\times10^5$ mock-null-hypothesis datasets were generated, and for the Milliquas catalog $7\times10^4$ datasets.
The red dashed vertical line indicates the test statistic for the neutrino doublet coincident with NGC 7469.
}\label{fig:ts_distr}
\end{figure}

\
To infer a p-value from the outcome of the test statistic in Sect.~\ref{subsec:teststatistic}, it is necessary to know its distribution under the hypothesis of alerts that did not originate from any source in the catalog.
By generating pseudo-random right ascensions for the set of neutrino alerts described in Sect.~\ref{subsec:realtime} we created mock-null-hypothesis data.
By finding for each mock-set of alerts the doublet with the highest test statistic, we are able to explore the test-statistic distribution under the $H_0$ hypothesis.
Figure~\ref{fig:ts_distr} shows the distribution of the test statistic under the $H_0$ hypothesis for the two considered catalogs.
Table~\ref{table:results} reports the results from the various Goodness-of-Fit tests.
Appendix~\ref{sec:decvar} studies the impact of a variation in the declination of the alerts in the mock uncorrelated dataset and finds a negligible impact.

Considering $2$ trials for testing the two catalogs, the global p-value is $8.0\times10^{-4}$, equivalent to a Gaussian one-sided significance of $3.16\,\sigma$.
A similar value is obtained when using the intrinsic X-ray flux as a tracer or assigning equal weights to all sources (see Appendices~\ref{sec:xraytest} and~\ref{sec:noweight}).

\begin{deluxetable*}{c c c c c}\label{table:results} 
\tabletypesize{\small}
\tablewidth{0pt} 
\tablecaption{Goodness-of-Fit tests results}
\tablehead{
Catalog & Best-doublet & Source & p-value & p-value (in $\sigma$)
}
\colnumbers
\startdata 
Milliquas & IC220424A \& IC230416A & NGC 7469 & $4.5\times10^{-3}$ & 2.61 \\
Turin     & IC220424A \& IC230416A & NGC 7469 & $4.0\times10^{-4}$ & 3.35 \\
\enddata
\end{deluxetable*}

\section{Discussion}\label{sec:discussion}
\
We exclude the null hypothesis for the two neutrinos in coincidence with NGC\,7469 at the level of 3.16~$\sigma$.
This result leaves open the possibility that either one or both of the neutrinos originated from the source.

In addition to the SplineMPE reconstruction, we apply, as a consistency check, our test to the Millipede reconstructions. Given the much larger area of the uncertainty areas of the Millipede algorithm, we do not expect to pick up a significant result, and thus we do not count this as an extra trial. We rather test if our analysis picks up the same doublet and source, which it did (see Appendix~\ref{sec:mill_gof} for more details).

\begin{figure*}
\plotone{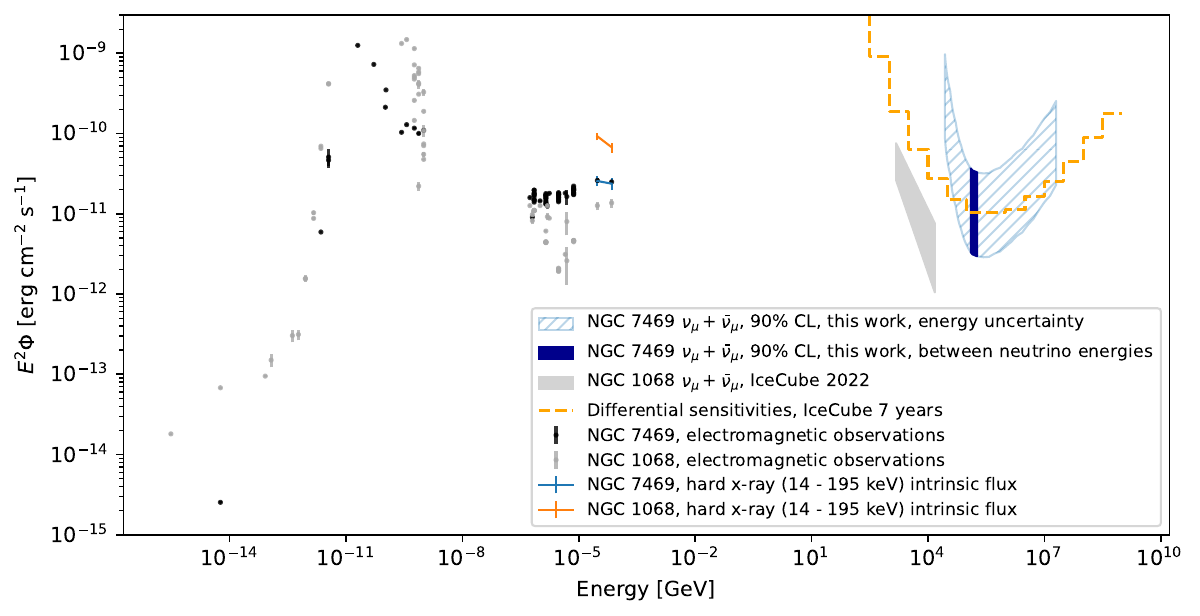}
\caption{Multimessenger SEDs of the two sources NGC 7469 and NGC 1068. Electromagnetic observations from \citet{swbat70, 4XMM-DR12, XMMSL2, RXS2CAT, 1SXPS, allwise, planck, AT20G, 2mass, VLSSr, NVSS}. Neutrino flux of NGC 1068 from \citet{ngc1068neutrinos}. Differential sensitivities from \citet{7ypstracks}.
Intrinsic X-ray fluxes from \citet{BASSxray}.
The confidence interval for the emission of NGC\,7469 was estimated in this work. The width in energy of the confidence interval spans from 27\,TeV to 20\,PeV, reflecting the uncertainty on the true neutrino energy. The shape of the confidence interval reflects the dependence of IceCube's effective area for realtime alerts reported in \cite{abbasi2023icecat1} and has nothing to do with the neutrino energy spectrum of the source.}
\label{fig:sed}
\end{figure*}

In the following, we estimate the possible neutrino flux from NGC\,7469 under different assumptions.
First, although our test does not rule out the possibility that only one neutrino originated from the source, we assume that both neutrinos (IC220424A and IC230416A) are indeed emitted from the source, since we consider this to be the most interesting case. Second, we assume a steady neutrino emission over the full length of our sample, i.e. 4 years of operations of the realtime system \citep{Blaufuss:20199c}. 
Two neutrinos are not sufficient to estimate a spectral index. 
We use the information that 2 neutrinos in 4 years were detected from the source to estimate a 90\% confidence interval on the neutrino rate $\lambda_{\nu_\mu+\bar{\nu}_\mu}$, independent of the energy.
This estimation was based on Poisson statistics.
We then convert these upper and lower limits on the neutrino rate $\lambda_{\nu_\mu+\bar{\nu}_\mu}$ into a confidence interval on the neutrino energy flux $\Phi_{\nu_\mu+\bar{\nu}_\mu}$.
This has a strong dependence on the energy:
\begin{equation}\label{eq:nueflux}
    \Phi_{\nu_\mu+\bar{\nu}_\mu} = \frac{\lambda_{\nu_\mu+\bar{\nu}_\mu}}{E A_\mathrm{eff}(E)} \, ,
\end{equation}
where $E$ is the neutrino energy and $A_\mathrm{eff}(E)$ is IceCube's effective area at that energy.
The GCN Notices sent out by IceCube report the most likely neutrino energy, assuming an astrophysical neutrino flux described by a power law with spectral index $-2.19$ \citep{Blaufuss:20199c}.
This is not a direct measure of the neutrino energy, and it is important to consider the respective uncertainties. Since uncertainties are not reported for the individual events, we perform a rough estimate following the example of the through-going neutrino event IC170922A~\citep{txs2018}, which was found in coincidence with the blazar TXS 0506+056. 
\citet{txs2018} reports a  most likely neutrino energy of 290\,TeV and a lower limit at 90\% confidence level (CL) of 183\,TeV. 
We include as well an upper limit at 20 PeV, indicative of the highest energies that IceCube should reasonably be able to detect.
We adopt here a lower limit of 27\,TeV (100\,TeV less than the lowest energy of the two neutrino events) and an upper limit of 20\,PeV. 
We note that this range is only used for visualization of the uncertainty on the energy range in Fig.~\ref{fig:sed} (shaded blue band), where we re-estimate the upper and lower limits on the neutrino energy flux for the various energies using Eq.~\ref{eq:nueflux}. 
The range between the reported neutrino energies is displayed as a blue band.
We also provide an estimate of the 90\% confidence interval of the neutrino flux at 161\,TeV (the average energy of the two neutrinos): $\Phi_{\nu_\mu+\bar{\nu}_\mu}=(0.68,7.56)\times10^{-16}$ TeV$^{-1}$cm$^{-2}$s$^{-1}$.
An alternative flux estimation considering all the alerts included in the IceCat catalog~\citep{abbasi2023icecat1} is presented in Appendix~\ref{sec:12yrs}.

This flux is compared to the differential sensitivity (dashed orange line in Fig.~\ref{fig:sed}) from an all-sky search for time-integrated neutrino emission from astrophysical sources with
7 years of IceCube data \citep{7ypstracks}.
The expected flux from NGC 7469 is right at the limit of this differential sensitivity.
The 7-year analysis, nor any precedent IceCube works with archival data \citep{ngc1068neutrinos, 7ypstracks, 10yrspstracks}, revealed an excess of neutrinos from this source. 
If NGC\,7469 is indeed a neutrino emitter, the lack of previous hints of neutrino emission from this source might be twofold: (i) the neutrino spectrum might be different from a soft power law, might be either a power law with a hard spectral index or a non-power-law spectrum peaked at high energies; (ii) the neutrino emission is variable with time and increased in the last years, which were not covered by previous analyses based on data recorded prior to the detection of the two neutrino alerts. 
A hard neutrino spectrum could be explained by magnetized strongly turbulent corona \citep{Fiorillo:2024akm, kohtahardspectrum}.

Electromagnetic observations of NGC 1068 and NGC 7469 are shown alongside the neutrino fluxes in Fig~\ref{fig:sed} for comparison. The estimated flux of NGC 7469 lies at higher energies and is larger compared to the extrapolation of the power-law flux from NGC 1068. The estimated intrinsic X-ray flux is lower, but harder, in the case of NGC\,7469. Possibly higher-energy photons could be the target for $p\gamma$ interactions in NGC\,7469.

Detailed modelling of the particle interaction and radiation processes, which are outside the scope of this paper, will help to evaluate the conditions required for efficient neutrino production in NGC\,7469.
\\
\\
\textit{We acknowledge support from the
Deutsche Forschungsgemeinschaft through the Collaborative Research Center SFB 1491 “Cosmic
Interacting Matters - from Source to Signal.”}

\newpage
\appendix

\section{Goodness of Fit test using X-ray fluxes as weights}\label{sec:xraytest}
As several models suggest the intrinsic X-ray flux as a good tracer for neutrino emission \citep{Inouexray,kotaxray,alixray}, we performed an additional test using the intrinsic X-ray fluxes in the 14-195 keV energy band estimated by \citet{BASSxray} to weight the Seyfert galaxies in the Turin-SyCAT catalog \citep{turincatalog}.
67 sources in the Turin-SyCAT catalog \citep{turincatalog} are not present in \citet{BASSxray}.
However, these 67 sources are all dim in observed X-rays when compared to NGC 7469 or NGC 1068.
Therefore, we decided to exclude these sources from the test.
To weight with the X-ray flux instead of the redshift, we modified eq.~\ref{eq:expectednumnu} to:

\begin{equation}\label{eq:expectedmuxray}
    \mu\left(\theta_S, \phi_{x,S}\right) = C_x\phi_{x,S}TE_0^2\sum_kA_\mathrm{eff}\left(\theta_S,E_k\right)\frac{E_k^+ - E_k^-}{E_k^+ E_k^-}\,,
\end{equation}

where $\phi_{x, S}$ is the X-ray energy flux in the energy band 14-195 keV from \citet{BASSxray} and $C_x$ is a proportionality factor between $\phi_{x, S}$ and $\phi_0$, the neutrino rate at the energy $E_0$, where $E_0$ is the lower bound in the effective-area energy range from \citep{abbasi2023icecat1}.
We assume $C_x$ being the same for all sources. 
We choose a small $C_x$ to make the test independent on its specific choice, as explained for $\phi_0$ in sec.~\ref{subsubsec:signal} and shown in Fig.~\ref{fig:flux}. 
More precisely, we choose $C_x < 10^{-3}$~GeV$^{-2}$.
As in the test with the redshift, we assume that all sources emit neutrinos following a power-law spectral shape with spectral index $\gamma = -2$.

We find a p-value of $2.4\times10^{-4}$, equivalent to 3.49~$\sigma$ (without any trial correction).
Also in this test, IC220424A and IC230416A together with NGC 7469 gave the best test statistic.
We performed the test with the reconstruction from the first GCN Notice (SplineMPE). For completeness, we repeated the test with the reconstruction from the updated GCN Circular (Millipede), see Appendix~\ref{sec:mill_gof}. 
The result is more significant than the same test performed using the redshift as weight (Sec.~\ref{sec:results}). 
Nevertheless, we keep considering the test with the redshift as our main result for two main reasons.  First, the redshift is a measurement much less model dependent than the intrinsic X-ray flux and represents a more agnostic choice. 
Second, because of the lack of estimations for the intrinsic X-ray fluxes, we had to exclude 67 sources (19$\%$ of all sources in the catalog) from our catalog. 
Even if these 67 sources were all dim in observed X-rays when compared to NGC 7469 and NGC 1068, this might not hold for their respective intrinsic X-ray fluxes. 

\section{Goodness of fit test using equal weights}\label{sec:noweight}
The choice to weight the sources using the redshift~(Sec.~\ref{subsubsec:signal}) was made to be as agnostic as possible.
An alternative agnostic approach would be to weight all sources in the same way.
However, assigning equal weights to the sources inevitably places greater importance on the source selection.
The 71345 X-ray-detected AGNs from the Milliquas catalog~\citep{milliquas} correspond to an average density of~$\sim1.7$ sources per square degree.
A neutrino alert with a $90\%$ uncertainty radius of $0.6$ degrees would contain an average of $\sim2.0$ sources within its contour.
With the redshift weighting, this was not a major issue since most AGNs were heavily penalized due to their distance.
However, when all sources are given equal weight, the concept of ``chance coincidence with a source'' is lost, as each neutrino will have multiple significant coincidence.
For this reason, we apply this test with equal weighting only to the Turin catalog~\citep{turincatalog}.

To weight the sources equally, we modified eq.~\ref{eq:expectednumnu} to:

\begin{equation}\label{eq:expectedmunoweight}
    \mu\left(\theta_S\right) = \phi_{\mathrm{equal}}TE_0^2\sum_kA_\mathrm{eff}\left(\theta_S,E_k\right)\frac{E_k^+ - E_k^-}{E_k^+ E_k^-}\,,
\end{equation}

where $\phi_\mathrm{equal}$ is the flux normalization which is the same for each source and has no dependence on the redshift.
This system still preserves the information about the effective area of IceCube.
As in Sec.~\ref{subsec:choices} and Appendix~\ref{sec:xraytest}, we still operate in the regime of low fluxes to keep the outcome of the test independent of the specific choice of the flux normalization by choosing $\phi_\mathrm{equal}<10^{-7}$~s$^{-1}$GeV$^{-1}$cm$^{-2}$.

We find a p-value of $5.9\times10^{-4}$, equivalent to $3.24$~$\sigma$ (without any trial correction).
In this test, IC220424A and IC230416A together with NGC~7469 gave the best test statistic.
We performed the test with the reconstruction from the first GCN~Notice (SplineMPE).
For completeness, we repeated the test with the reconstruction from the updated GCN Circular (Millipede); see Appendix~\ref{sec:mill_gof}.
We do not consider this test as the main one, as it entails a strong dependence on the source selection and is not applicable to the Milliquas catalog.
However, we include it in this appendix as proof of the consistency of the main result and its robustness, despite using different agnostic approaches.

\section{Goodness of Fit test using Millipede errors}\label{sec:mill_gof}

For completeness, we performed the same goodness-of-fit test of Sec.~\ref{sec:test}, Appendix~\ref{sec:xraytest} and Appendix~\ref{sec:noweight} using the uncertainties from the GCN Circulars (computed using Millipede, see Sec.~\ref{sec:neutrinodata}). We do not expect this test to be sensitive given the large uncertainty contours.
Also in this case, we used the Gold and Bronze alerts released since 2019, until the 4 October 2023. 
This set of alerts includes one additional event, IC210503A \citep{2021GCN.29951....1I}, which had no automated GCN Notice and therefore was not used for the test with SplineMPE (Sec.~\ref{sec:neutrinodata}).

In the GCN Circulars, the 90\% uncertainty region is reported as a rectangle. 
For our statistical test, we need to translate this rectangle into the $\sigma$ of a bivariate Gaussian (Sec.~\ref{sec:test}).
We retrieve this information by performing the following steps:
\begin{enumerate}
    \item Calculate the angular area $A_{90}$ for the rectangular uncertainty region in the GCN circulars; 
    \item Find the 90$\%$ error radius $R_{90}$ for the circle with an area $A_{\mathrm{Circle}} = A_{90}$;
    \item Scale $R_{90}$ to the $\sigma$ of the bivariate Gaussian distribution, assuming it as the point-spread function\footnote{
    This is a strong approximation and summary of the information content in the rectangular region. 
    Nevertheless, the test statistic needs just a parameter to order the various coincidences. 
    The $\sigma$ is to be understood solely for this purpose.}.
\end{enumerate}
The resulting p-values are the following:
\begin{itemize}
    \item Milliquas catalog and redshift weighting: $0.26$ (0.64~$\sigma$);
    \item Turin catalog and redshift weighting: $3.7\times10^{-2}$ (1.79~$\sigma$);
    \item Turin catalog and X-ray flux weighting: $1.4\times10^{-2}$ (2.20~$\sigma$);
    \item Turin catalog and no weighting: $2.3\times10^{-2}$ (1.99~$\sigma$).
\end{itemize}
None of the p-values is significant.
Nevertheless, IC220424A and IC230416A together with NGC 7469 always return the highest test statistic, proving a consistency of our tests.
However, as expected, the p-values obtained using the smaller contours of SplineMPE are much more significant.
This difference of the results using the two reconstruction can be explained by the sizes of the uncertainty areas.
An algorithm with a higher precision can significantly improve the sensitivity of an experiment.

\section{Sensitivity of the test to the injection of alerts from sources}\label{sec:unbias}

In the Goodness-of-Fit (GoF) test a null and an alternative hypothesis are used to build the test statistic (Sec.~\ref{sec:test}). 
However, the test is designed to validate or reject only the null hypothesis, and not to make any conclusion on the alternative one. 
This characteristic gives the GoF test some freedom on the specific choice of the alternative hypothesis, which does not necessarily have to fully describe the (unknown) reality.
Under the alternative hypothesis chosen for the test, the neutrino doublet with the highest test statistic should be the most likely one to be produced by a source.
Here, we apply a sanity check by injecting neutrino doublets from sources in the scrambled neutrino datasets~(Sec.~\ref{sec:results}) to study how the test statistic distribution varies.
Because the alternative hypothesis assumes that the neutrino doublet has been produced by a source~(Sec.~\ref{subsec:teststatistic}), the test statistic distribution should shift towards higher test-statistic values.

Moreover, we study how the test statistic distribution changes if only one neutrino (a singlet) is injected at a source position.
In this case, we expect the distribution to shift to higher values, but less compared to the case of a doublet injection.

\begin{figure}
\plotone{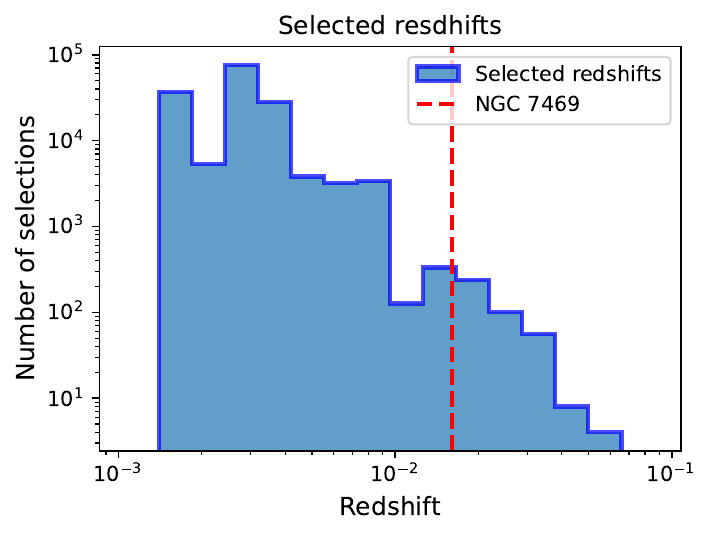}
\caption{
Redshift of the sources selected from the Turin catalog~\citep{turincatalog} for the doublet and singlet injection. The red dashed vertical line indicates the redshift of NGC~7469.
}\label{fig:redshift_distr}
\end{figure}

\begin{figure}
\plotone{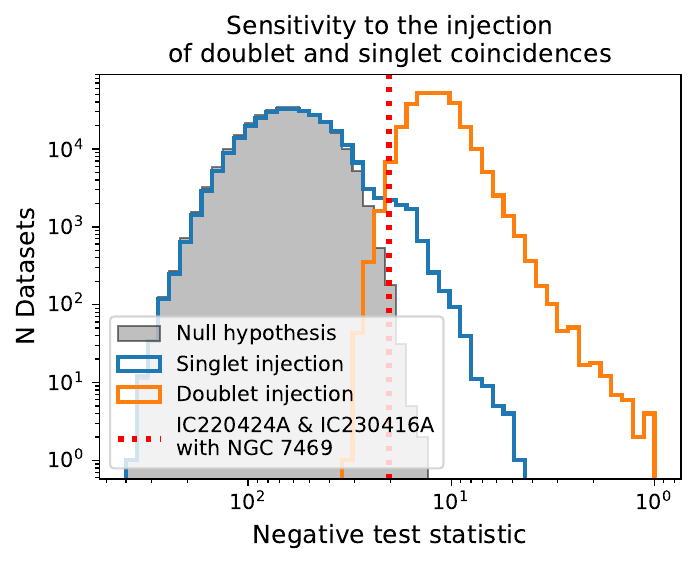}
\caption{
Test statistic distributions for scrambled datasets with neutrinos that did not originate from any source in the catalog (in grey), one neutrino injected at a source position (in blue), and a doublet injected at a source position (in orange).
$3\times10^5$ mock-datasets were generated for the null-hypothesis and alternative scenarios. For the latter two, two and one random alerts of each mock dataset are shifted on top of one source from the catalog, selected randomly according to the weighting scheme. This method keeps the total number of alerts constant.
The red dotted vertical line indicates the test statistic for the neutrino doublet coincident with NGC 7469.
For all distributions, the uncertainty from the first GCN Notice (SplineMPE) with the Turin catalog \citep{turincatalog} were used.
}\label{fig:unbiasedness}
\end{figure}

To verify these behaviors, we generated mock datasets with one signal doublet (or singlet) each, which we placed on the position of one source in our catalog. The signal doublet (singlet) is generated by shifting two (one) alerts from the mock sample to the position of the source.
Each time, the catalog source of the doublet (singlet) is selected randomly according to the ordering given by the probability in eq.~\ref{eq:pf}.
Figure~\ref{fig:redshift_distr} shows the redshifts of the sources from the Turin catalog~\citep{turincatalog} selected with this system.
NGC~7469 is not one of the nearest sources, but it is also not totally unlikely for a source at a similar redshift to be selected.
The directions of the two (one) neutrinos of the mock doublet (singlet) were generated following the spatial distribution in eq.~\ref{eq:spatialcomp} and using two (one) spatial uncertainties randomly selected from the uncertainties of the neutrino dataset.
The resulting distributions with the doublet and singlet injections, compared to the distribution under the null hypothesis (from Sec.~\ref{sec:results}), are shown in Fig.~\ref{fig:unbiasedness} for the SplineMPE neutrino dataset and the Turin catalog~\citep{turincatalog}.

The test-statistic distribution with the doublet-from-source injection correctly shifts towards higher values.
On the other hand, the distribution with the singlet injection is mostly similar to the null-hypothesis distribution, except for a tail at the highest test-statistic values.
This tail corresponds to cases where one scrambled neutrino accidentally overlaps with the same source where an alert was injected.
This scenario is more likely than having two neutrinos randomly overlapping a source.
However, since the test statistic is designed to recognize doublets coincident with a source and not singlets, our test is not very efficient at distinguishing between the case of one neutrino correlated with a source and the null-hypothesis case.

Regarding the doublet injection, the observation of the neutrino doublet of IC220424A and IC230416A coincident with NGC~7469 is more compatible to its distribution compared to the distribution under the null hypothesis (Fig.~\ref{fig:unbiasedness}).
The test is sensitive to the scenario simulated, although the alternative hypothesis does not fully describe that case, since it assigns a neutrino flux to each source that is low and makes the production of a doublet unlikely (Sec.~\ref{subsubsec:signal}).
An additional simplification lies in the assumption of a power-law neutrino spectral shape with spectral index of $\gamma=2$ for all sources.

The lack of knowledge on the neutrino production mechanism and its tracers makes us refrain from validating a given model prediction. 

\section{Estimation of the neutrino flux of NGC~7469 using the IceCat catalog of neutrino alerts}\label{sec:12yrs}

After the last update of the IceCube Realtime System \citep{Blaufuss:20199c}, IceCube reprocessed its data to look for events which would have passed the criteria for realtime alerts.
The results of this reprocessing are publicly available as the IceCat catalog \citep{abbasi2023icecat1}, which contains events starting from 2011.
However, this catalog contains the per-event directional information of the GCN Circular (Millipede) only. For this reason, the additional years have not been used in this work.
Nevertheless, it can still be used to check if other neutrinos in the past came from the same direction.

One additional neutrino event in the IceCat catalog \citep{abbasi2023icecat1}, IC190619A, has a directional estimation compatible with the position of NGC 7469.
However, IC190619A is also present in the dataset of neutrino events (gold and bronze alerts starting from 2019) used in this work, but with NGC~7469 outside the 90\% contour of the first GCN~Notice.
Therefore, this event was not considered as contributing to the neutrino emission.
To remain compatible with the rest of this work, we do not consider IC190619A as related to NGC 7469 for the flux estimate.
Accordingly, by looking at all 12 years (from 2011 to 2023) covered by the IceCat catalog, we find no further events coincident with the Seyfert galaxy besides the doublet of IC220424A and IC230416A.

\begin{figure}
\plotone{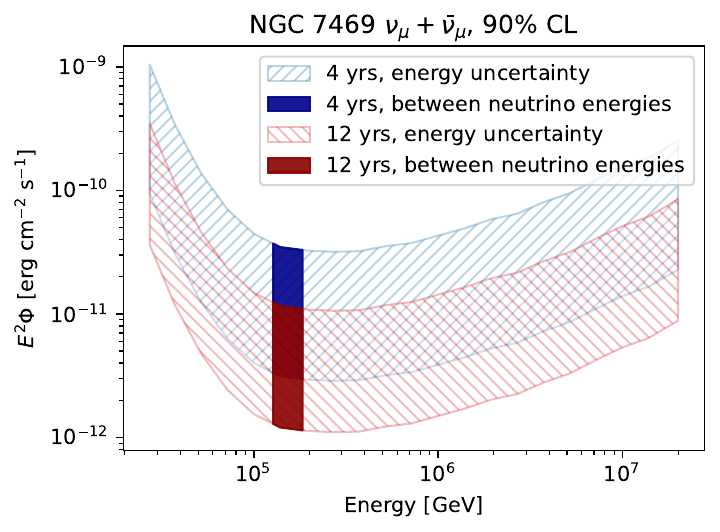}
\caption{
Confidence intervals for the neutrino emission of NGC 7469 estimated using 4 years of data (in blue, corresponding to the gold and bronze alerts released by IceCube since 2019), and 12 years of data (in red, corresponding to the neutrino events in the IceCat catalog). 
The width in energy of the confidence intervals spans from 27 TeV to 20 PeV, reflecting the uncertainty on the true neutrino energy.
The shape of the confidence intervals reflects the dependence of IceCube’s effective area for realtime alerts reported in \citet{abbasi2023icecat1} and is not related to the neutrino energy spectrum of the source.
}\label{fig:12yrsflux}
\end{figure}

We repeated the same estimation of the possible neutrino flux from NGC 7469 as in Sec.~\ref{sec:discussion} this time using the information that 2 neutrinos in 12 years were detected.
Figure~\ref{fig:12yrsflux} shows the result of this estimation, compared to the result from Sec.~\ref{sec:discussion}.
We also provide a new estimate of the 90\% confidence interval of the neutrino flux at 161\,TeV (the average energy of the two neutrinos): $\Phi_{\nu_\mu+\bar{\nu}_\mu}=(0.26,2.52)\times10^{-16}$ TeV$^{-1}$cm$^{-2}$s$^{-1}$.

\section{Declination variation in the generation of mock-null-hypothesis neutrino data}\label{sec:decvar}
In Sec.~\ref{sec:results}, to evaluate the test-statistic distribution under the null hypothesis, only the right-ascension of the neutrino alerts was varied in the generation of the mock-null-hypothesis data.
The declinations were kept unchanged, as the effective areas of IceCube depends on the declination.
To verify that small variations in the declination do not affect the outcome of our test, we repeat the generation of the mock-null-hypothesis data by adding a small random variation to the declination of the single alert. We chose the variation according to a uniform distribution between $-x$ and $x$, with $x$ of 1, 2, and 3 degrees.
With $x=1$~deg the resulting p-value is $5.3\times10^{-4}$ (3.27~$\sigma$). With $x=2$~deg it is $7.6\times10^{-4}$ (3.17~$\sigma$). With $x=3$~deg it is $7.9\times10^{-4}$ (3.16~$\sigma$). When comparing these results with the p-value from the original GoF test ($4.0\times10^{-4}$, equivalent to 3.35~$\sigma$) we can conclude that our results are not strongly affected by the small shift.



\bibliography{bibliography}{}
\bibliographystyle{aasjournal}



\end{document}